\documentclass[3p]{elsarticle}

\usepackage{newtxmath,newtxtext}
\usepackage{mathtools}
\usepackage{tikz}
\usepackage[range-phrase=\text{--},range-units=single]{siunitx}

\usetikzlibrary{patterns}
\usetikzlibrary{fadings}

\newcommand\rossby\varepsilon
\newcommand\disturb\alpha

\newcommand\dt[2]{\frac{\partial^{#2} #1}{\partial t^{#2}}}
\newcommand\dx[2]{\frac{\partial^{#2} #1}{\partial x^{#2}}}
\newcommand\dy[2]{\frac{\partial^{#2} #1}{\partial y^{#2}}}
\newcommand\dz[2]{\frac{\partial^{#2} #1}{\partial z^{#2}}}

\newcommand\inftml{\mathrm{d}}
\newcommand\I{\mathrm{i}}
\newcommand\e{\mathrm{e}}
\newcommand\diff[3]{\frac{\mathrm{d}^{#3} #1}{\mathrm{d} #2^{#3}}}
\newcommand{\reynolds}{Re}
\newcommand\dotprod{\boldsymbol{\cdot}}

\renewcommand\vec[1]{\boldsymbol{#1}}

\begin{document}

\title{Stability of SQG Kolmogorov Flow}

\author[1]{Mac Lee}
\ead{mal004@ucsd.edu}
\author[2,3]{Stefan G.\ Llewellyn Smith}
\ead{sgls@ucsd.edu}

\affiliation[1]{Department of Physics, University of California, San Diego CA
92093-0319, USA}
\affiliation[2]{Department of Mechanical and Aerospace Engineering, Jacobs
School of Engineering, University of California, San Diego CA 92093-0411, USA}
\affiliation[3]{Scripps Institution of Oceanography, University of California,
San Diego CA 92093-0209, USA}

\begin{abstract}
  Stability analysis is performed on surface quasigeostrophic systems subjected
  to a Kolmogorov-type ``shear force'' on the boundaries using linear and
  nonlinear approaches.
  For a SQG system of semi-infinite depth forced on the upper boundary, the most
  linearly unstable mode is $2.74$ the energy injection length scale.
  This is contrary to two-dimensional fluid systems, where the linear
  instability is greatest for long waves.
  In the presence of damping, the most linearly unstable mode shifts toward
  shorter length scales.
  The nonlinear critical Reynolds number across different damping strengths is
  found to be qualitatively similar to that of Euler 2D systems.
  For an SQG system of finite thickness being forced on both boundaries, its
  behaviour approaches that of a semi-infinite SQG system at the large thickness
  limit.
  In the small thickness limit, the behaviour of a symmetrically forced fluid
  layer approaches that of a 2D system, while an antisymmetrically forced fluid
  layer is not susceptible to both linear and nonlinear instabilities.
  With ageostrophic effects, an SQG+ system of semi-infinite depth is much more
  prone to instabilities than an otherwise identical SQG system in the absence
  of damping due to the instability of long-wave modes.
  However, damping significantly suppresses such instabilities.
  With increasing damping, the most
  linearly unstable mode moves toward a smaller length scale.
  Contrary to the zero damping case, when the damping is sufficiently large,
  ageostrophic effects have a small but measurable stabilising effect.
\end{abstract}


\maketitle

\section{Introduction}

First proposed by~\cite{blumen1978}, SQG theory was suggested by a few
studies~\citep{held1995,lapeyre2006,lacasce2006} to be a possible model to
relate observable sea surface height to internal dynamics of the upper ocean in
the submesoscale. \cite{lapeyre2006} argued that the balanced flow resulting
from SQG strongly anticorrelated with the balanced flow from QG in the upper
ocean, and the overall dynamics is dominated by the SQG component. However, in
the mid-latitudes, the Rossby number is not vanishingly small in the
submesoscale, and SQG theory does not account for the presence of ageostrophic
dynamics in this length scale. Since SQG theory consists of the first order
terms of a perturbation expansion in the Rossby number, it is useful to consider
next order correction terms to capture such dynamics. Such correction terms for
QG were derived by~\cite{muraki1999}. In Muraki's theory, the QG streamfunction
is replaced by a vector potential, with the components governed by three Poisson
equations. An ansatz given in a follow-up study which applied the same idea to
SQG effectively transforms the next order equations for SQG from Poisson
equations to Laplace equations~\citep{hakim2002}. Inverting these equations, one
can derive diagnostic expressions of the correction terms in terms of the SQG
streamfunction.

To cast light on the physical properties of SQG dynamics, the natural first step
is to study known, exactly solvable configurations. Kolmogorov flow is one such
configuration. In Kolmogorov flows, a viscous fluid is forced by a sinusoidal
body force at one length scale, and this injection of energy is balanced by
dissipation. This is one of the simplest settings in which different flow
properties can be explored. Two-dimensional Kolmogorov flow has been
successfully realised in laboratory
experiments~\citep{bondarenko1979,burgess1999}. Theoretically, the stability of
this class of problem has been studied using various
techniques~\citep{meshalkin1961,beaumont1981,gotoh1983,sivashinsky1985,gotoh1987}.
The usual first step is to study the linearised stability problem. The critical
Reynolds number for such a system, computed using the continued fraction method,
is $\sqrt{2}$~\citep{meshalkin1961}. Further studies have been carried out on
the effects of the $\beta$-plane~\citep{manfroi2002,tsang2008}, Ekman
drag~\citep{fukuta1995,tsang2008}, and buoyancy~\citep{balmforth2002}. However,
linear analysis is only valid when the flow is slightly disturbed from the basic
state. Even then, it only provides an upper bound to stability: just because a
configuration has a Reynolds number below the predicted critical Reynolds number
does not guarantee the stability of the configuration. To get a complementary
picture of the stability properties, the energy method can be used with the
nonlinear equation to obtain a lower bound for the critical Reynolds number,
below which the system is guaranteed to be
stable~\citep{fukuta1995,straughan2004,tsang2008}. There have been studies of
SQG Kolmogorov flows using Galerkin methods~\citep{kalashnik2020,kalashnik2022},
so that the stability of a heavily truncated system is considered, but an
analysis of SQG Kolmogorov flow stability for linear and energy stability does
not appear to have been carried out. With these tools, we are able to reuse the
machinery developed for two-dimensional Kolmogorov flow in prior studies to
study order Rossby effects in SQG+.

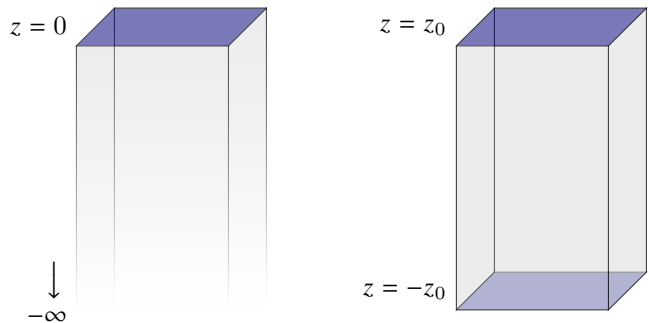
\begin{figure}
  \begin{center}
    \pgfdeclarelayer{foreground}
    \pgfsetlayers{main,foreground}
    \begin{tikzpicture}
      \begin{scope}
        \node[left] at (-0.5,-0.25) {$z=0$};
        \node[below,rotate=-90] at (-0.6,-3.6) {$\longrightarrow$};
        \node[left] at (-0.5,-4.1) {$-\infty$};
        \begin{pgfonlayer}{foreground}
          \draw[fill=blue!50!black,semitransparent] (0,0) coordinate (A)
            -- ++(2,0) coordinate (B)
            -- ++(-0.5, -0.5) coordinate (C)
            -- ++(-2, 0) coordinate (D)
            -- cycle;
        \end{pgfonlayer}
        \draw[fill=black!10,semitransparent,path fading=south] (A) -- (B) -- ++(0,-3.5) -- ++(-2,0) -- cycle;
        \draw[fill=black!10,semitransparent,path fading=south] (B) -- (C) -- ++(0,-3.5) -- ++(0.5,0) -- cycle;
        \draw[fill=black!10,semitransparent,path fading=south] (C) -- (D) -- ++(0,-3.5) -- ++(2,0) -- cycle;
        \draw[fill=black!10,semitransparent,path fading=south] (D) -- (A) -- ++(0,-4) -- ++(-0.5,0) -- cycle;
      \end{scope}
      \begin{scope}[shift={(5,0)}]
        \node[left] at (-0.5,-0.25) {$z=z_0$};
        \node[left] at (-0.5,-3.75) {$z=-z_0$};
        \begin{pgfonlayer}{foreground}
          \draw[fill=blue!50!black,semitransparent] (0,0) coordinate (A)
            -- ++(2,0) coordinate (B)
            -- ++(-0.5, -0.5) coordinate (C)
            -- ++(-2, 0) coordinate (D)
            -- cycle;
        \end{pgfonlayer}
        \draw[fill=blue!50!black,semitransparent] (0,-3.5) coordinate (E)
          -- ++(2,0) coordinate (F)
          -- ++(-0.5, -0.5) coordinate (G)
          -- ++(-2, 0) coordinate (H)
          -- cycle;
        \draw[fill=black!10,semitransparent] (A) -- (B) -- (F) -- (E) -- cycle;
        \draw[fill=black!10,semitransparent] (B) -- (F) -- (G) -- (C) -- cycle;
        \draw[fill=black!10,semitransparent] (C) -- (G) -- (H) -- (D) -- cycle;
        \draw[fill=black!10,semitransparent] (D) -- (H) -- (E) -- (A) -- cycle;
      \end{scope}
    \end{tikzpicture}
  \end{center}
  \caption{Illustration of the chosen system configurations. The
  first figure is semi-infinite, the second has finite  depth.}
  \label{fig:illustration-of-configuration}
\end{figure}

In this study, we choose to focus on two specific configurations. The first
configuration is a semi-infinite fluid layer driven at the upper boundary. This
configuration is a very simple idealisation of the upper ocean, where the upper
boundary is the air-sea interface. The second configuration is a finite fluid
layer that is driven at both the upper and lower boundaries with fluid below it. This is a more
detailed model of the upper ocean, where the mixed layer is bounded on top by
the sea surface and on the bottom by the thermocline. The mixed layer is forced
on both boundaries by air-sea interaction on top and interactions between the
mixed layer and the stratified layer below. These configurations are illustrated in
Figure~\ref{fig:illustration-of-configuration}.

In the following sections, we carry out stability analyses using
linearisation and the energy method for the SQG problem. \S~\ref{sec:formulation} gives a brief discussion of the SQG formulation of
the Kolmogorov problem. In \S~\ref{sec:semi-infinite-layer-linear-instability},
we study the linear instability of a semi-infinite system with a constant
Brunt-V\"ais\"al\"a frequency driven at the upper boundary. This is followed by
a study of the linear instability of a finite-depth system in
\S\,\ref{sec:finite-depth-layer-linear-instability}. The behaviours of both
symmetric and antisymmetric forcing are studied. In
\S\,\ref{sec:semi-infinite-layer-nonlinear-stability} and
\S\,\ref{sec:finite-depth-layer-nonlinear-stability}, we study the nonlinear
stability problems of the same configurations using the energy stability method. In
\S\,\ref{sec:linear-instability-order-rossby}, we include $O(\epsilon)$ corrections in the linear stability analysis or the semi-infinite configuration. \S\,\ref{sec:discussion} concludes the paper.

\section{Formulation of the Problem}\label{sec:formulation}

The governing equations for an SQG system on the upper boundary of a fluids occuping $z < 0$
are similar to those of a two-dimensional fluid system, with the relation between streamfunction and the active scalar changing.
An SQG system for the upper and lower boundaries of  fluid layer with $z \in [-z_0, z_0]$ should behave like a semi-infinite SQG system
in the limit of large fluid layer thickness, and like a two-dimensional system
in the small thickness limit. This is because the effect of boundary forcing
decreases exponentially with distance, so that as the thickness of the fluid is
increased, the dynamics of the fluid close to the boundaries will be
increasingly dominated by the closest boundary, effectively turning the system
into two decoupled semi-infinite SQG systems.
Conversely, if the forcing on the two boundaries is symmetric, the forcings from
both boundaries act in concert, and their combined effect in the fluid interior
becomes increasingly less dependent on $z$ as the thickness is reduced,
effectively reducing the system into a 2D system.

The active scalar in SQG is the surface potential temperature
$\theta = \partial\psi/\partial z$, where $\psi$ is the streamfunction. The time
evolution of the tracer is governed by the nondimensional
advection equation of $\theta$~\citep{fukuta1995,muraki1999}:
\begin{equation}
  \dt{\theta}{} + \vec{u}\dotprod\boldsymbol{\nabla}\theta
  = -\frac{1}{\reynolds_n} [(-1)^{n}\laplace^n\theta+\lambda\theta] + F,
  \label{eq:fullAdvEq}
\end{equation}
where $\reynolds_n = UL^{2n-1}/\nu$ is the Reynolds number; 
$n$ is the hyperdiffusion index, which governs the scales at which diffusion
takes effect ($n=1$ is regular viscosity), $\lambda$ is linear drag or
(radiative) damping scaled with the Reynolds number,
$\laplace\equiv\nabla^2=\partial^2/\partial x^2+\partial^2/\partial y^2$ is the
horizontal Laplacian, and $F$ is a source/sink term. In Kolmogorov
flow, the forcing $F(y, z)$ is a periodic function of $y$. There is a
laminar basic state with a balance between the energy injection at the boundary
and dissipation. In SQG, the internal potential vorticity is zero,
\begin{equation}
  \nabla_3^2 \psi = 0,
\end{equation}
where $\nabla_3^2 = \partial^2/\partial x^2+\partial^2/\partial
y^2+\partial^2/\partial z^2$ is the three-dimensional Laplacian.

\section{SQG Linear Instability}

\subsection{Semi-infinite depth fluid}
\label{sec:semi-infinite-layer-linear-instability}


The laminar stationary solution with vanishing potential vorticity is
\begin{subequations}
  \label{eq:basicState}
  \begin{equation}
    \label{eq:basicPsi}
    \Psi(y,z) = \Psi_0\e^z\sin y,
  \end{equation}
  from which we obtain the potential temperature function of the basic state,
  \begin{equation}
    \label{eq:basicTheta}
    \Theta(y,z) = \Psi_0\e^z\sin y.
  \end{equation}
\end{subequations}
This corresponds to the forcing
\begin{equation}
  \label{eq:kolForcing}
  F
    = \frac{\Psi_0}{\reynolds_n}\left(\lambda + 1\right)\e^z\sin y.
\end{equation}

To examine linear disturbances to this basic state, we write 
\begin{equation}
  \psi = \Psi + \tilde{\psi} \qquad \theta = \Theta + \tilde{\theta},
\end{equation}
where $\tilde{\psi}$ and $\tilde{\theta}$ are disturbance terms. This leads to
the exact equation on the surface
\begin{equation}
  \frac{\partial\tilde{\theta}}{\partial t} + J(\tilde{\psi},\tilde{\theta}) +
  \Psi_0(\tilde{\theta} + \tilde{\psi})_x\cos y
  = -\frac{1}{\reynolds_n}\{(-1)^{n}\laplace^{n}+\lambda\}\tilde{\theta},\label{eq:pertAdv}
\end{equation}
where $f_x = \partial f/\partial x$ and the Jacobian is given by $J(f,g) = f_x g_y - f_y g_x$.
To obtain the growth rate of the unstable states, we
substitute a Floquet ansatz
\begin{equation}
  \tilde{\psi}(x,y,z,t) = \e ^{\sigma t + \I kx + \I ly}\sum_{m =
  -\infty}^{\infty}\psi_m \e ^{\I my + \kappa_m z},\label{eq:floquet_ansatz_semi-infinite}
\end{equation}
into \eqref{eq:pertAdv}, writing $\kappa_n = \sqrt{k^2 + (l + n)^2}$. When the disturbance is
small, the nonlinear term $J(\tilde{\phi}, \tilde{\theta})$ in
\eqref{eq:pertAdv} can be neglected. Substituting these equations into the
linearised advection equation for $\theta$ at the upper boundary,
we obtain
\begin{equation}
  \label{eq:single_interface_o1_recurrence_relation}
  \frac{\I k\Psi_0}{2}\frac{\kappa_{m+1} - 1}{\kappa_m}\psi_{m + 1}
  - \frac{1}{\reynolds_n}\left(\kappa_m^{2n}+\lambda\right)\psi_m
  + \frac{\I k\Psi_0}{2}\frac{\kappa_{m-1} - 1}{\kappa_m}\psi_{m - 1}
  = \sigma\psi_m.
\end{equation}
This infinite set of coupled equations is formally an infinite matrix eigenvalue problem. It can be solved numerically by truncating the sum to $-N \leq
m \leq N$. In  tests, we found no noticeable difference between $N=64$ and $N=16$. For the remainder of the paper, $N$ is $16$ unless otherwise
specified. At $\lambda=0$, with $\Psi_0=1$ and $n=1$, we find that the
critical Reynolds number is $4.64671$ numerically, corresponding to $l = 0$.

\begin{figure}
  \begin{center}
    \includegraphics[width=2.6in]{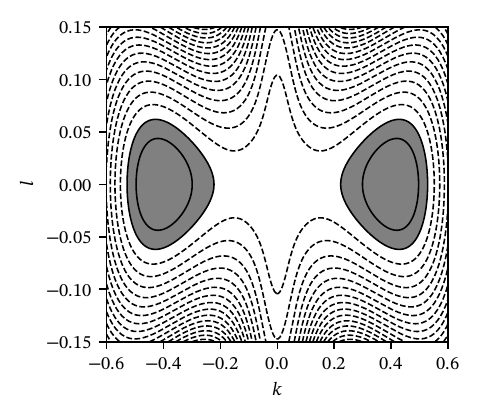}
    \includegraphics[width=2.6in]{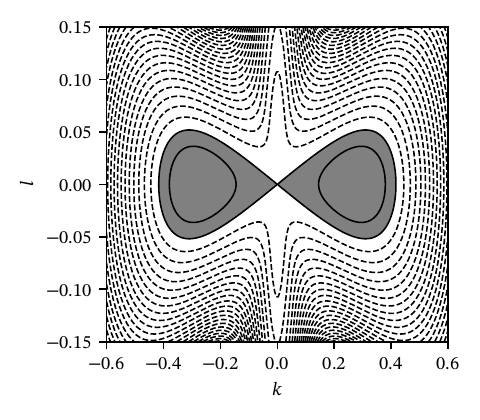}
    \caption{Left: real part of the growth rate at $\reynolds_n = 5.02$. Dotted
    contours indicate a negative growth rate $\sigma$, solid contours indicate a
    positive growth rate. Right: analogous plot for a two-dimensional system at
    $\reynolds_n = 1.83$. The Reynolds numbers are chosen sp that the regions of instability are
    equal in size. In both plots, $n=1$ and
    $\lambda=0$.}\label{fig:single_interface_sqg}
  \end{center}
\end{figure}

\begin{figure}
  \begin{center}
    \includegraphics[width=2.6in]{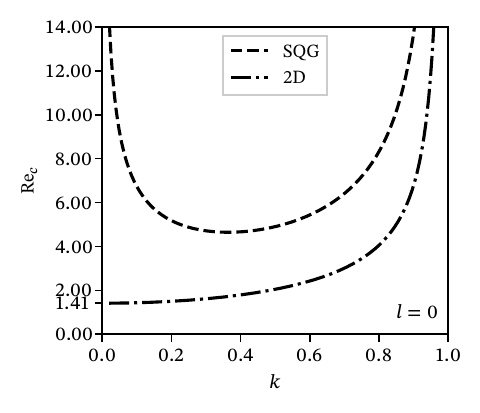}
    \includegraphics[width=2.6in]{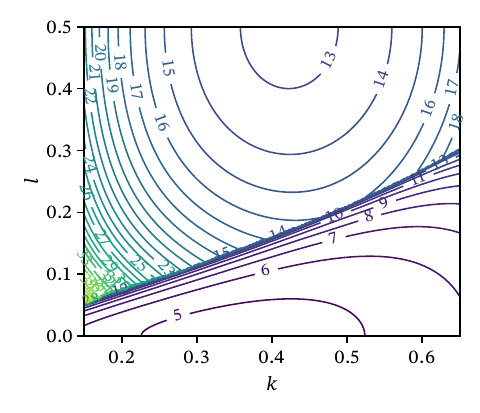}
    \caption{Left: critical Reynolds number as a function of $k$ and $l = 0$ with $n=1$ and $\lambda=0$, correspondingg to the smallest critical Reynolds number. Right: critical Reynolds
    number as a function of the wavenumbers $k$ and
    $l$.}\label{fig:k_v_critical_reynolds}
  \end{center}
\end{figure}

\begin{figure}
  \begin{center}
    \includegraphics[width=2.6in]{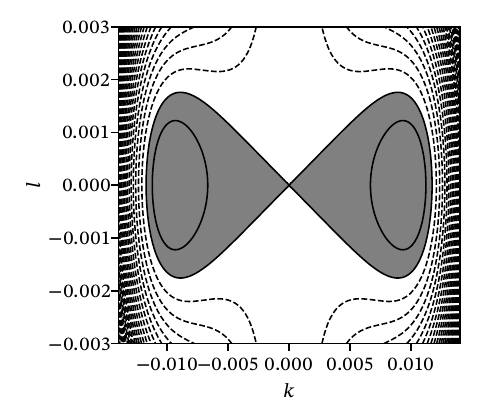}
    \includegraphics[width=2.6in]{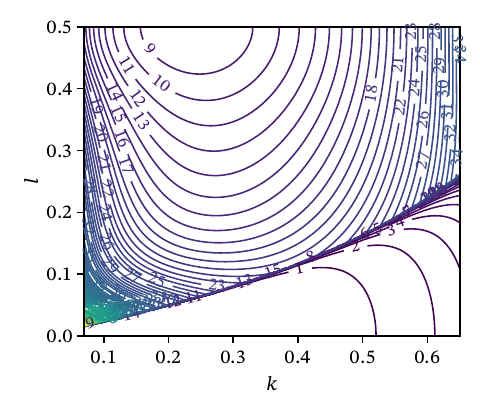}
  \end{center}
  \caption{Left panel: growth rate for hyperdiffusion parameter
  $n=4$ with Reynolds number $\reynolds_n=\num{3e-5}$; $\sigma>0$ in the grey region. Right panel: contour map of critical Reynolds
  numbers for $n=4$ and $\lambda=0$.}
  \label{fig:semi-infinite-hyperdiffusion}
\end{figure}


Figure~\ref{fig:single_interface_sqg} shows that, at $\reynolds_1 = 5.02$, taken
to be an illustrative value showing the form of the instability region. A region
of instability emerges at line $l = 0$ near $k = 0.36$. The region of
instability is disconnected from $(k,l)=(0,0)$. This is  different from a
similar plot of a two-dimensional system where the instability occurs at $k = l
= 0$ in the long-wave limit and $\reynolds_1 = \sqrt{2}$. This can be further
illustrated by plotting the critical Reynolds number as a function of $k$ along
$l=0$, as shown in Figure~\ref{fig:k_v_critical_reynolds}. In a two-dimensional
system, with ordinary diffusion, the minimum critical Reynolds number
corresponds to the mode at $k=l=0$, whereas the critical Reynolds number is
infinite at the same point in a semi-infinite SQG system. This indicates that in
two-dimensional systems, modes with large length scale are the least stable,
whereas such modes are linearly stable in SQG systems. In both systems, the
critical Reynolds number approaches infinity at $k=1$, and the modes at $k>1$
are linearly stable. Modes smaller than the length scale of energy injection are
linearly stable in both Euler 2D and semi-infinite SQG. It is also helpful to
visualise the critical Reynolds number as a contour plot in the horizontal wave
numbers $k$ and $l$, as  in Figure~\ref{fig:k_v_critical_reynolds}.
These plots show that local interactions play a more important role in SQG
dynamics than in 2D dynamics.

As the hyperdiffusive parameter $n$ increases to $4$, the region of instability
in wavenumber space once again connects to the origin of the plane, as shown in
Figure~\ref{fig:semi-infinite-hyperdiffusion}. Compared to regular diffusion,
hyperdiffusion more selectively suppresses small scale wave modes, enabling the
large scale wave modes to once again determine the linear instability of the
system. This mechanism is also much less efficient at removing energy, which
manifested in a dramatic drop in the critical Reynolds number down to
$\reynolds_c=\num{2.01055e-5}$.


\begin{figure}
  \begin{center}
    \includegraphics[width=\textwidth]{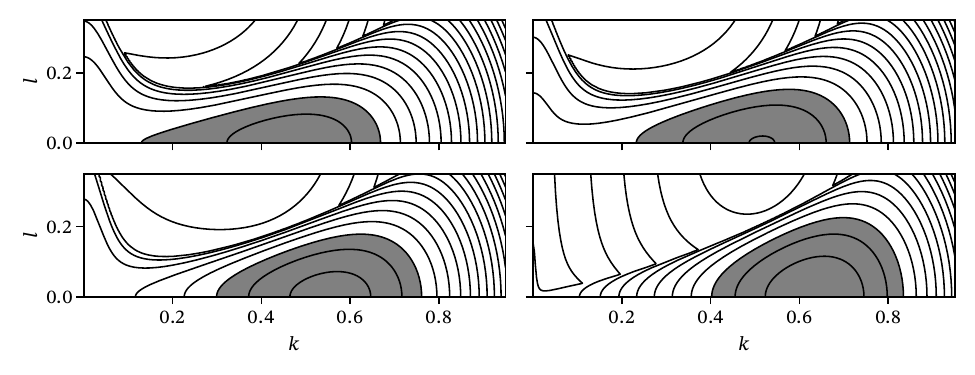}
    \caption{Growth rate diagrams for semi-infinite SQG
    at $n=1$ and $1.3\reynolds_c(\lambda)$. From left to right, top to bottom,
    $\lambda$ is $0$, $0.05$, $0.2$, $5$. The contour spacing is $0.01$.}
    \label{fig:single_interface_growth_rate_varied_lambda_above_neutral_line}
  \end{center}
\end{figure}

\begin{figure}
  \begin{center}
    \includegraphics[width=2.6in]{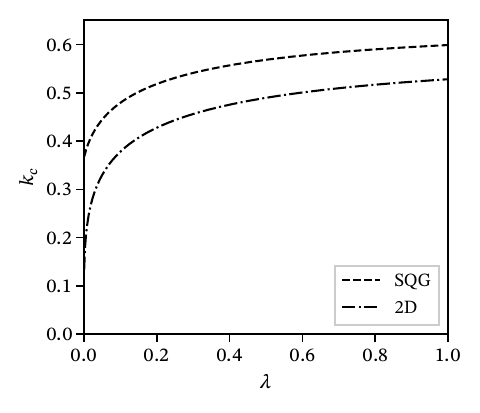}
  \end{center}
  \caption{Plot of the critical wavenumber $k_c$ as a function of damping
  $\lambda$. A plot of a two-dimensional curve is shown for comparison. $k_c$
  saturates at $0.650$ in SQG.}
  \label{fig:single_interface_linear_o1_kc_v_lambda}
\end{figure}

\begin{figure}
  \begin{center}
    \includegraphics[width=2.6in]{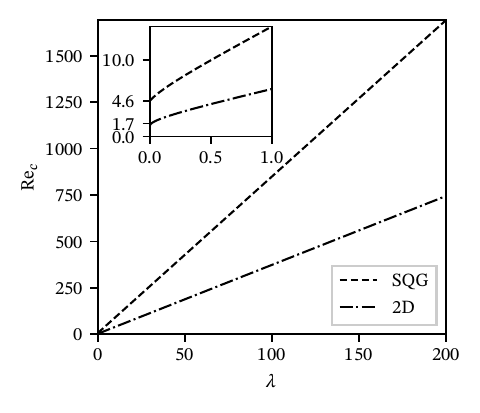}
    \includegraphics[width=2.6in]{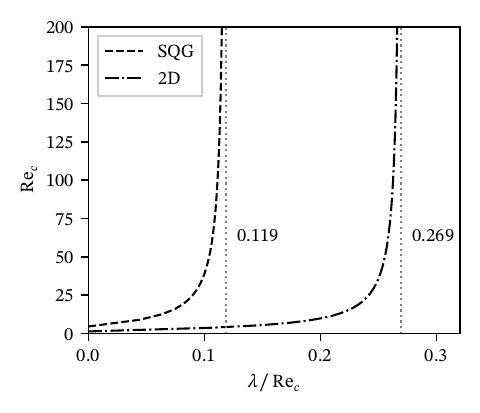}
  \end{center}
  \caption{The first plot shows the critical Reynolds number as a function of
  the damping parameter $\lambda$. The second plot shows the critical Reynolds
  number as function of physical damping $\lambda/\reynolds_c$.}
  \label{fig:single_interface_linear_Re_c_v_lambda}
\end{figure}

Figure~\ref{fig:single_interface_growth_rate_varied_lambda_above_neutral_line}
shows the growth rate diagram with increasing $\lambda$, with the Reynolds
numbers just above the critical number at the given damping. The critical
Reynolds number increases with damping. This relationship is shown in
Figure~\ref{fig:single_interface_linear_o1_kc_v_lambda}. Since damping removes
energy from the system across length scales, increasing damping decreases the
linear instability of the system. The region of instability shifts to higher $k$
values as $\lambda$ is increased. The regions of instability also get more
stretched in the $l$ direction. This relationship is further illustrated with a
plot of the most unstable wave mode $k_c$ as a function of $\lambda$, as shown
in Figure~\ref{fig:single_interface_linear_o1_kc_v_lambda}. Without damping, the
point at $k_c=0.364$ is the most linearly unstable mode. As damping increases,
this point shifts toward larger wavenumbers and saturates at around $k_c=0.650$.
The curve for a two-dimensional system is shown for comparison. Physical damping
is defined by $\lambda/\reynolds$ with respect to the notation in
\eqref{eq:fullAdvEq}. As physical damping increases, the removal of energy
exceeds the injection of energy beyond a certain threshold, and the system can
no longer become linearly unstable. In other words, the linear critical Reynolds
number becomes infinity. This is shown graphically in
Figure~\ref{fig:single_interface_linear_Re_c_v_lambda}.
For an SQG system, this threshold is
at $\lambda/\reynolds_c=0.119$, which is considerably lower than that of an
otherwise identical Euler 2D system. The most unstable wave mode cannot move
beyond $k_c=0.650$ with increasing physical damping because the system ceases to
be linearly unstable beyond that point. In short, SQG Kolmogorov systems are
less linearly unstable than 2D systems, and the linear instabilities also tend
to be associated with smaller length-scales.

\subsection{Boundary driven
finite-depth fluid}\label{sec:finite-depth-layer-linear-instability}


In a model driven on two boundaries by Kolmogorov type forcing, the basic state
can be trivially generalised from the basic state in the previous section. For
mathematical simplicity, we place the two boundaries at $z=\pm z_0$. The
configuration is visualised by the second illustration in
Figure~\ref{fig:illustration-of-configuration}. The basic state is then a linear
combination of two basic states corresponding to the two boundaries. We study
the stability of such a system by looking at the symmetric and antisymmetric
configurations separately.

We start by taking a look at the symmetric configuration. Adding the basic
states on the two boundaries, the symmetric basic state in the bulk between the
two boundaries is
\begin{subequations}
  \label{eq:two_boundaries_basic_symm}
  \begin{align}
    \Psi(y, z)
               &= 2\Psi_0 \e ^{z_0}\cosh z\sin y,\label{eq:two_boundaries_basic_psi_symm} \\
    \Theta(y, z) &= 2\Psi_0\e ^{z_0}\sinh z\sin y.\label{eq:two_boundaries_basic_theta_symm}
  \end{align}
\end{subequations}
This basic state is maintained by the forcing at the boundary,
\begin{equation}
  F = \frac{2\Psi_0}{\reynolds_n}\left(\lambda + 1\right)\e ^{-z_0}\sinh z\sin y.
\end{equation}
On the boundaries, we require that $w=0$. Using
\eqref{eq:two_boundaries_basic_symm}, the advection
equation \eqref{eq:pertAdv} becomes
\begin{equation}\label{eq:sqg_twolayer_perturb}
  \dt{\tilde{\theta}}{} + 2\Psi_0 \e ^{-z_0}\left(\tilde{\theta}\cosh z_0 \pm \tilde{\psi}\sinh z_0\right)_x\cos y  = -\frac{1}{\reynolds_n}\left\{(-1)^{n}\nabla^{2n} + \lambda\right\}\tilde{\theta}
\end{equation}
at $z = \pm z_0$ after linearisation. Using the Floquet ansatz
\begin{equation}
  \tilde{\psi}(x, y, z, t) = \e ^{\I kx+\I ly+\sigma t} \sum_{m=-\infty}^{\infty} \psi_m(z) \e ^{\I my},
\end{equation}
\eqref{eq:sqg_twolayer_perturb} becomes
\begin{multline}\label{eq:sqg_twolayer_perturb_floquet}
  \left\{\sigma + \kappa_m^{2n} + \lambda\right\} \psi_m(z) \\
  =\I k\Psi_0\e ^{-z_0} \left\{[\psi_{m+1}(z) + \psi_{m-1}(z)] \cosh z_0 \mp [\psi_{m+1}(z) \pm \psi_{m-1}(z)] \sinh z_0\right\}.
\end{multline}
By definition, the QG potential vorticity is zero in the fluid interior in SQG.
The streamfunction must be a solution of Laplace's equation.
Consequently, the $z$-dependence must assume the form
\begin{equation}\label{eq:sqg_twolayer_z_dep}
  \psi_m(z) = \mu_m \cosh\kappa_m z + \nu_m \sinh\kappa_m z,
\end{equation}
where $\mu_m$ and $\nu_m$ are coefficients to be determined.
\begin{subequations}
  Combining  \eqref{eq:sqg_twolayer_perturb_floquet} and
 ~\eqref{eq:sqg_twolayer_z_dep} and evaluating at $z=\pm z_0$,
  we obtain the recurrence relation for the symmetric modes,
  \begin{multline}
    \sigma {\mu}_m = -\frac{\I k \Psi_0 \e ^{- z_0} (\kappa_{m + 1} \sinh \kappa_{m + 1} z_0 \cosh z_0 - \sinh z_0 \cosh \kappa_{m + 1} z_0)}{\kappa_m \sinh z_0 \kappa_m} {\mu}_{m + 1}\\
    - \frac{\I k \Psi_0 \e ^{- z_0} (\kappa_{m - 1} \sinh \kappa_{m - 1} z_0 \cosh z_0 - \sinh z_0  \cosh \kappa_{m - 1} z_0)}{\kappa_m \sinh z_0 \kappa_m} {\mu}_{m - 1} \\
    - \frac{\lambda + \kappa_m^{2 n}}{\reynolds_n} {\mu}_m.
    \label{eq:double_boundary_symmetric_A}
  \end{multline}
  Similarly, the recurrence relation for the antisymmetric modes is
  \begin{multline}
    \sigma {\nu}_m = -\frac{\I k \Psi_0 \e ^{-z_0} (\kappa_{m + 1} \cosh z_0  \cosh z_0 \kappa_{m + 1} - \sinh z_0  \sinh \kappa_{m + 1} z_0)}{\kappa_m \cosh z_0 \kappa_m} {\nu}_{m + 1} \\
    - \frac{\I k \Psi_0 \e ^{-z_0} (\kappa_{m - 1} \cosh z_0  \cosh \kappa_{m - 1} z_0 - \sinh z_0 \sinh \kappa_{m - 1} z_0)}{\kappa_m \cosh z_0 \kappa_m} {\nu}_{m - 1} \\
    - \frac{\lambda + \kappa_m^{2 n}}{\reynolds_n} {\nu}_m.
    \label{eq:double_boundary_symmetric_B}
  \end{multline}
\end{subequations}

For the antisymmetric configuration, the streamfunction and surface potential
temperature of the basic state are
\begin{subequations}
  \begin{align}
    \Psi(y, z) &= 2\Psi_0 \e ^{z_0}\sinh z\sin y\label{eq:two_boundaries_basic_psi_antisymm} \\
    \Theta(y, z) &= 2\Psi_0\e ^{z_0}\cosh z\sin y.\label{eq:two_boundaries_basic_theta_antisymm}
  \end{align}
\end{subequations}
Taking similar steps as for the symmetric configuration, the recurrence relations
for symmetric and antisymmetric modes are
\begin{multline}
  \sigma {\mu}_m = - \frac{\I k \Psi_0 \e ^{-z_0} (\kappa_{m + 1} \sinh z_0  \cosh \kappa_{m + 1} z_0 - \sinh \kappa_{m + 1} z_0 \cosh z_0)}{\kappa_m \sinh \kappa_m z_0} {\nu_0}_{m + 1} \\
  - \frac{\I k \Psi_0 \e ^{-z_0} (\kappa_{m - 1} \sinh z_0  \cosh \kappa_{m-1} z_0  - \sinh \kappa_{m - 1} z_0  \cosh z_0)}{\kappa_m \sinh z_0 \kappa_m} {\nu_0}_{m - 1} \\
  - \frac{\lambda + \kappa_m^{2 n} {\mu}_m}{\reynolds_n},
\end{multline}
\begin{multline}
  \sigma {\nu}_m = - \frac{\I k \Psi_0 \e ^{-z_0} (\kappa_{m + 1} \sinh z_0  \sinh \kappa_{m + 1} z_0 - \cosh z_0  \cosh\kappa_{m + 1}z_0)}{\kappa_m \cosh z_0 \kappa_m} {\mu_0}_{m+1} \\
  - \frac{\I k \Psi_0 \e ^{-z_0} (\kappa_{m - 1} \sinh z_0 \sinh \kappa_{m - 1}z_0 - \cosh z_0 \cosh \kappa_{m - 1} z_0)}{\kappa_m \cosh z_0 \kappa_m} {\mu_0}_{m - 1} \\
  - \frac{\lambda + \kappa_m^{2 n}}{\reynolds_n} {\nu}_m.
\end{multline}

\begin{figure}
  \begin{center}
    \includegraphics[width=2.6in]{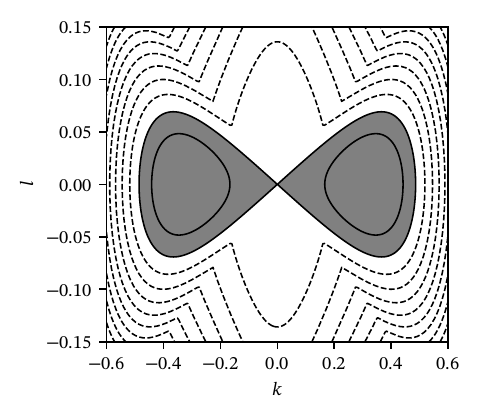}
    \includegraphics[width=2.6in]{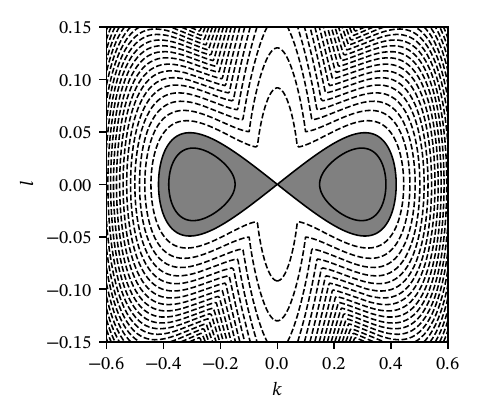}
    \caption{Left panel: real part of the growth rate of the
    linearised system for a symmetric configuration, similar to
    Figure~\ref{fig:single_interface_sqg}. The growth rate is positive in the
    grey region. This plot corresponds to $z_0=2.2$ and $\reynolds_n=3.8$.
    Right: the growth rate plot for the antisymmetric
    configuration with identical parameters. The unstable region in $k$-$l$
    space is enlarged with the symmetric configuration compared to the
    antisymmetric configuration.
    }\label{fig:vt}
  \end{center}
\end{figure}

\begin{figure}
  \begin{center}
    \includegraphics[width=2.6in]{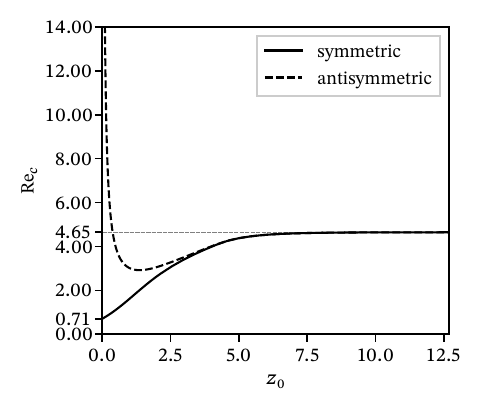}
    \includegraphics[width=2.6in]{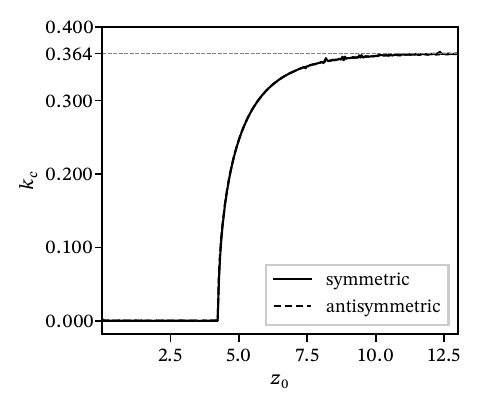}
    \caption{Left panel: critical Reynolds number $\reynolds_c$ vs.~$z_0$. The horizontal line indicates $\reynolds_n=4.6467$. The critical
    Reynolds number for symmetric forcing increases monotonically until it
    saturates at around $\reynolds_n=4.6467$. With antisymmetric forcing,
    the critical Reynolds number is infinite at $z_0=0$ and then rapidly decreases.
    The neutral line of antisymmetric forcing asymptotically converges to that of
    symmetric forcing. Right panel: the corresponding horizontal wavenumber $k$.
    At low separation distances, the
    wavenumber corresponding to the most unstable mode $k_c$ is identically
    zero,  as it is in a two-dimensional system. At a separation distance larger than
    4.22, $k_c$ moves outward on the $k$-axis and approaches $k_c=0.3640$,
    which corresponds to $k_c$ in semi-infinite SQG.
    }\label{fig:double_boundary_z0_vs_Re_c}
  \end{center}
\end{figure}

\begin{figure}
  \begin{center}
    \includegraphics[width=\textwidth]{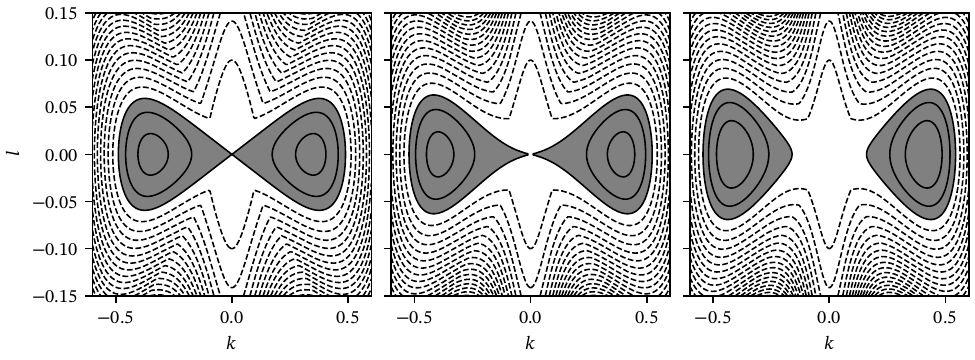}
    \caption{Growth rate diagrams at $z_0$ at 4, 6, 8 and $\reynolds_n$ at 4.67,
    4.978 and 5.092 for symmetric forcing. The Reynolds numbers were chosen to
    produce diagrams with regions of instability with the same size.
    $\lambda=0$, $n=1$, $N=16$. At small separation distances, the growth rate
    diagram resembles that of a two-dimensional system; at large separation
    distances, the growth rate diagram resembles that of an SQG system with a
    semi-infinite layer.}
    \label{fig:two-active-boundaries-symmetric-o1-vary_z0}
  \end{center}
\end{figure}

\begin{figure}
  \begin{center}
    \includegraphics[width=0.5\textwidth]{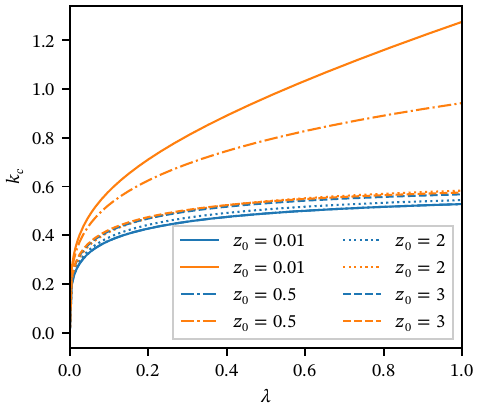}
  \end{center}
  \caption{Critical wavenumber $k_c$ as a function of damping
  $\lambda$ at various separation distances. The orange (blue) curves corresponds
  to the (anti)symmetric configuration.}
  \label{fig:double_boundary_kc_vs_lambda}
\end{figure}

The growth rate diagrams for both the symmetrically and antisymmetrically forced
system are shown in Figure~\ref{fig:vt}. The region of instability of the
symmetrically forced system is noticeably bigger than the antisymmetrically
forced system for the chosen $z_0$ and $\reynolds_n$. This behaviour is due to
the critical Reynolds number being slightly lower in the symmetric configuration
than the antisymmetric configuration at this separation distance $z_0$. The
relationship of the critical Reynolds number with $z_0$ is shown in
Figure~\ref{fig:double_boundary_z0_vs_Re_c}. When $z_0 \gg 1$, critical Reynolds
number asymptotically approaches the value obtained for SQG in the semi-infinite
fluid layer in \S~\ref{sec:semi-infinite-layer-linear-instability}. When $z_0
\ll 1$, the critical Reynolds numbers of these two configurations diverge, with
$\reynolds_c$ becoming infinity for the antisymmetric configuration at $z_0=0$.
The critical Reynolds number for the symmetric configuration becomes
$1/\sqrt{2}$ when $z_0$ approaches $0$. The specific effect of layer thickness
in a boundary driven finite-thickness SQG system on the growth rate diagram ca
be seen in Figure~\ref{fig:two-active-boundaries-symmetric-o1-vary_z0}. At low
$z_0$, the regions have a teardrop shape and are connected to the origin of the
$k$-$l$ plane. The regions of instability are connected to the origin even when
the Reynolds number is only slightly above $\reynolds_c$, just as it is in a
two-dimensional system. As $z_0$ is increased, the teardrop-shaped regions of
instability retreat toward higher $|k|$ when the Reynolds number is slightly
above $\reynolds_c$, indicating the point in wavenumber space that first gets
unstable shifts to larger $|k|$ as the fluid layer thickness is increased. The
panel on the right resembles that of the semi-infinite SQG system. This
behaviour is further illustrated by plotting the critical wavenumber $k_c$
against $z_0$, as shown in the right panel of
Figure~\ref{fig:double_boundary_z0_vs_Re_c}. At small $z_0$, the most linearly
unstable state corresponds to the long-wave limit. This resembles the behaviour
of a two-dimensional system. However, as the layer thickness  is increased
beyond about $z_0=4.3$, the most unstable mode abruptly shifts away from the
two-dimensional value and stabilises at $k_c=0.364$, the semi-infinite SQG
value. This can be explained from the recurrence relation itself. At the large
separation limit, both \eqref{eq:double_boundary_symmetric_A} and
\eqref{eq:double_boundary_symmetric_B} reduce to
\begin{multline}
  \sigma_0 \e ^{\kappa_m z_0} {\Psi_0}_m = -\frac{\I k\Psi_0}{2}\frac{\kappa_{m+1}-1}{\kappa_m} \e ^{\kappa_{m+1}z_0}{\Psi_0}_{m+1}
  -\frac{\I k\Psi_0}{2}\frac{\kappa_{m-1}-1}{\kappa_m} \e ^{\kappa_{m-1}z_0}{\Psi_0}_{m-1} \\
  -\frac{\lambda+\kappa_m^{2n}}{\reynolds_n} \e ^{\kappa_m z_0} {\Psi_0}_m.
\end{multline}
When $k$ or $l$ is sufficiently big, $\kappa_m\sim\kappa_{m\pm1}$, and this
equation reduces to~\eqref{eq:single_interface_recurrence_relation}. We can
expect the small length scale modes to behave very similarly to that of a
semi-infinite SQG system.
At the small $z_0$ limit,~\eqref{eq:double_boundary_symmetric_A} reduces to
\begin{equation}
  \sigma {\mu}_m =
  -\frac{\I k\Psi_0 (\kappa_{m+1}^2 - 1)}{\kappa_m^2} {\mu}_{m+1}
  -\frac{\I k\Psi_0 (\kappa_{m-1}^2 - 1)}{\kappa_m^2} {\mu}_{m-1}
  -\frac{\lambda + \kappa_m^{2n}}{\reynolds_n} {\mu}_m.
\end{equation}
Comparing this with \eqref{eq:single_interface_o1_recurrence_relation}
confirms that the critical Reynolds number at this limit is $1/\sqrt{2}$.

\section{SQG Nonlinear Stability}

\subsection{The energy method}\label{sec:energy_method}

Linear stability analysis gives the critical Reynolds number above which small
disturbances grow. However, since the instability conditions are obtained under
the assumption of small perturbations of the basic state, the actual threshold
for instability could be much lower. To find an accurate picture of the line
across which stability is guaranteed, we need to take into account nonlinear
effects. We adapt the nonlinear stability analysis approach taken
by~\cite{fukuta1995} for SQG. The discussion is supplemented by
\cite{straughan2004} and \cite{tsang2008}. We start by noting that the kinetic
energy, as it is conventionally defined, can be written in terms of potential
temperature through Parseval's theorem if the horizontal boundaries are
periodic, with
\begin{equation}
  E \equiv -\frac12 \iint_S \psi\nabla^2\psi\:\inftml S
  = \frac12 \sum_{k=-\infty}^{\infty}\sum_{l=-\infty}^{\infty} \hat{\psi}_{k,l}|\kappa|^2\hat{\psi}_{-k,-l}
  \equiv \frac12 \iint_S \theta^2\:\inftml S,
\end{equation}
where $\kappa^2 = k^2 + l^2$. We define the disturbance energy as
\begin{equation}
  \label{eq:dist_energy}
  \mathcal{E} = \frac12 \iint_S \tilde{\theta}^2\:\inftml S.
\end{equation}
We focus on $n=1$ in the remainder of this section.
Using the advection equation, \eqref{eq:pertAdv}, the time evolution of disturbance energy is
\begin{align}
  \begin{split}
    \diff{\mathcal{E}}{t}{} &= \frac12 \diff{}{t}{} \iint_S \tilde{\theta}^2\:\inftml S
    = \iint_S \tilde{\theta} \dt{\tilde{\theta}}{}\:\inftml S \\
    &= \iint_S \tilde{\theta} \left\{\frac{1}{\reynolds}\left(
      \laplace-\lambda\right)\tilde{\theta}
      - J(\Psi, \Theta)
      - J(\Psi, \tilde{\theta})
      - J(\tilde{\psi}, \Theta)
      - J(\tilde{\psi}, \tilde{\theta})
    \right\}\:\inftml S.
  \end{split}
\end{align}
Since Kolmogorov flow is an equilibrium state, $J(\Psi, \Theta) = 0$. The
integrals of the two terms
$\tilde{\theta}J(\tilde{\psi},\tilde{\theta})$ and
$\tilde{\theta}J(\Psi,\tilde{\theta})$ can also be shown to be zero.
Therefore
\begin{equation}
  \diff{\mathcal{E}}{t}{} =
  \iint_S \tilde{\theta} \left\{\frac{1}{\reynolds}\left(
   \laplace - \lambda\right)\tilde{\theta}
      - J(\tilde{\psi}, \Theta)\right\}\:\inftml S.
\end{equation}
The equation can be rewritten as
\begin{equation}\label{eq:de_dt}
  \diff{\mathcal{E}}{t}{} = B - \frac{1}{\reynolds}D,
\end{equation}
where
\begin{equation}\label{eq:production_term}
  B \equiv \iint_S \tilde{\theta}J(\Theta, \tilde{\psi})\:\inftml S
\end{equation}
is the production term and
\begin{equation}
  D \equiv \iint_S\tilde{\theta}\left(\lambda - \laplace\right)\tilde{\theta}\:\inftml S
\end{equation}
is the dissipation term. It will be useful to define a critical Reynolds number,
\begin{equation}
  \frac{1}{\reynolds_c} = \max \frac{B}{D}.
\end{equation}
From this definition, the change of the disturbance energy over time is bounded by
\begin{equation}\label{eq:dedt_bound_d}
  \diff{\mathcal{E}}{t}{} \le D \left(\frac{1}{\reynolds_c} - \frac{1}{\reynolds}\right).
\end{equation}
Using the Poincar\'e inequality $ |\nabla \tilde{\theta}|^2 \ge C
|\tilde{\theta}|^2 $ where $C$ is a constant, and integrating by parts, we
obtain an upper bound for the dissipation term
\begin{equation}\label{eq:dissipation_upper_bound}
  D = \iint_S \tilde{\theta} \left(\lambda - \laplace\right)\tilde{\theta}\:\inftml S
  = \lambda\iint_S \tilde{\theta}^2\:\inftml S + \iint_S \left|\nabla\tilde{\theta}\right|^2\:\inftml S
  \ge 2 (\lambda + C) \mathcal{E}.
\end{equation}
With $\reynolds < \reynolds_c$, we can combine \eqref{eq:dedt_bound_d} and
\eqref{eq:dissipation_upper_bound} and obtain an upper bound for the rate of
change of energy,
\begin{equation}
  \label{eq:time_diff_ineq}
  \diff{\mathcal{E}}{t}{} \le 2 (\lambda + C) \left(\frac{1}{\reynolds_c} - \frac{1}{\reynolds}\right) \mathcal{E}.
\end{equation}
Using Gronwall's inequality, we can solve this equation and obtain an upper
bound for $\mathcal{E}$,
\begin{equation}
  \mathcal{E} \le \mathcal{E}_0\exp\left\{2 (\lambda + C) \left(\frac{1}{\reynolds_c} - \frac{1}{\reynolds}\right)t\right\}.
\end{equation}
In other words, the disturbance energy is guaranteed to monotonically decay if
$\reynolds < \reynolds_c$. The nonlinear critical Reynolds number can be
obtained by maximising $B/D$ by varying the disturbance streamfunction
$\tilde{\psi}$,
\begin{equation}
  \frac{\delta (B/D)}{\delta\tilde{\psi}} = 0,
\end{equation}
or equivalently,
\begin{equation}
  \delta B - \frac{1}{\reynolds} \delta D = 0.\label{eq:max_var}
\end{equation}
Since $\tilde{\psi}$ satisfies Laplace's equation, the functional variation
$\delta\tilde{\psi}$ must also satisfy Laplace's equation. Using
\eqref{eq:production_term}, we obtain
\begin{equation}
  \delta B = \iint_S \left\{\dz{\delta\tilde{\psi}}{}\ J(\tilde{\psi},\Theta) + \dz{\tilde{\psi}}{} J(\delta\tilde{\psi},\Theta) \right\}\:\inftml S.
\end{equation}
We integrate the first term by parts,
\begin{equation}
  \iint_S \dz{\delta\tilde{\psi}}{}J(\tilde{\psi},\Theta)\:\inftml S
  \equiv \I\iint_S \laplace^{1/2}(\delta\tilde{\psi})J(\tilde{\psi},\Theta)\:\inftml S
  = \I\iint_S\delta\tilde{\psi} \laplace^{1/2}J(\tilde{\psi},\Theta)\:\inftml S,
\end{equation}
since for functions $g$ and $h$ that are doubly periodic on the domain $S$, it
can be shown by Parseval's theorem that
\begin{equation}
  \iint_S g \laplace^{1/2}h\:\inftml S
  = \iint_S h\laplace^{1/2}g\:\inftml S.\label{eq:dz_int}
\end{equation}
Here $\laplace^{\alpha/2}$ is the fractional Laplacian defined by its
effect in $k$-space,
\begin{equation}
  \theta = (-\laplace)^{\alpha/2}\psi, \quad \hat{\theta}(\vec{k}) = |\kappa|^\alpha \hat{\psi}(\vec{k}),
\end{equation}
where $\kappa^2 = k^2 + l^2$. In SQG, $(-\laplace)^{1/2} \equiv
\partial/\partial z$. Integrating the second term by parts gives
\begin{subequations}
  \begin{align}
    \iint_S \dz{\tilde{\psi}}{} J(\delta\tilde{\psi},\Theta)\:\inftml S
    &= \iint_S\dz{\tilde{\psi}}{} \left(\dx{\delta\tilde{\psi}}{}\dy{\Theta}{}-\dy{\delta\tilde{\psi}}{}\dx{\Theta}{}\right)\:\inftml S \\
    &= -\iint_S \left\{\dx{}{}\left(\dz{\tilde{\psi}}{}\dy{\Theta}{}\right)-\dy{}{}\left(\dz{\tilde{\psi}}{}\dx{\Theta}{}\right)\right\} \delta\tilde{\psi}\:\inftml S \\
    &\equiv -\I\iint_S J(\laplace^{1/2}\tilde{\psi},\Theta) \delta\tilde{\psi}\:\inftml S.
  \end{align}
\end{subequations}
Therefore
\begin{equation}
  \delta B = \I\iint_S \left\{\laplace^{1/2}J(\tilde{\psi},\Theta) - J(\laplace^{1/2}\tilde{\psi},\Theta)\right\} \delta\tilde{\psi}\:\inftml S.\label{eq:deltaB}
\end{equation}
Likewise,
\begin{subequations}
  \begin{align}
    \delta D &= -\iint_S \left\{2\lambda\dz{\delta\psi}{}\dz{\tilde{\psi}}{}
    -\dz{\delta\tilde{\psi}}{}\laplace\tilde{\theta}
    -\tilde{\theta}\laplace\dz{\delta\tilde{\psi}}{}\right\}\:\inftml S \\
    &= \iint_S 2\left(\lambda-\laplace\right)\laplace\tilde{\psi}\delta\tilde{\psi}\:\inftml S\label{eq:deltaD}
  \end{align}
\end{subequations}
where we used~\eqref{eq:dz_int}. Combining  \eqref{eq:max_var},
\eqref{eq:deltaB} and \eqref{eq:deltaD}, we have
\begin{equation}
  \iint_S \left\{
  \frac{2}{\reynolds}(\lambda - \laplace)\laplace\tilde{\psi}
  - \I \laplace^{1/2}J(\tilde{\psi},\Theta)
  + \I J(\laplace^{1/2}\tilde{\psi},\Theta)\right\} \delta\tilde{\psi}\:\inftml S
  = 0.
\end{equation}
Since $\delta\tilde{\psi}$ is an arbitrary function, the expression in the brackets must be zero, so
\begin{equation}\label{eq:energy_eq}
  \frac{2}{\reynolds}\left(\lambda - \laplace\right)\laplace\tilde{\psi}
  - \I \laplace^{1/2}J(\tilde{\psi},\Theta)
  + \I J(\laplace^{1/2}\tilde{\psi},\Theta) = 0.
\end{equation}
To find the critical Reynolds number, we need to solve this eigenproblem.

\subsection{Semi-infinite depth
fluid}\label{sec:semi-infinite-layer-nonlinear-stability}

\begin{figure}
  \begin{center}
    \includegraphics[width=2.6in]{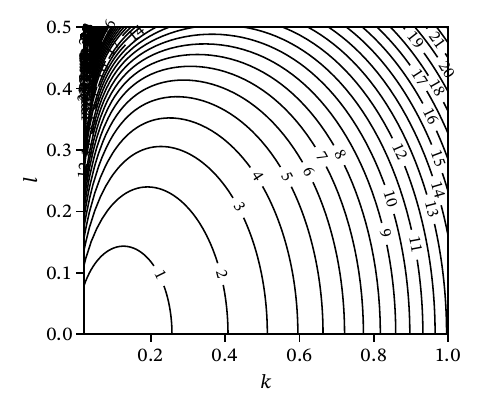}
    \includegraphics[width=2.6in]{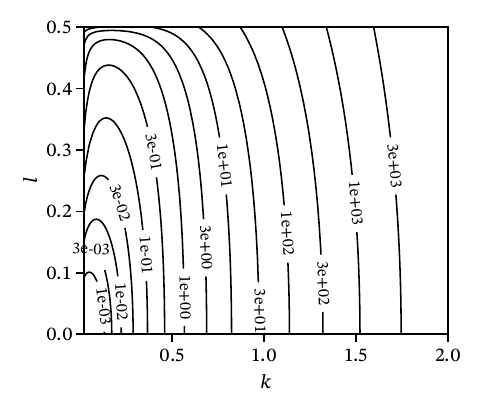}
    \caption{Critical Reynolds number using the energy method. Left:
    critical Reynolds number with hyperdiffusion parameter $n=1$. Right:
    critical Reynolds number with hyperdiffusion parameter $n=4$. $\lambda=0$ for
    both plots.}
    \label{fig:single_boundary_critical_reynolds_k_l}
  \end{center}
\end{figure}

For a semi-infinite fluid layer, we use the ansatz
\begin{equation}
  \tilde{\psi} = \e ^{\I kx + \I ly}\sum_{m=-\infty}^{\infty}\psi_m \e ^{\I my + \kappa_m z}.
\end{equation}
Substituting into the equation, one obtains an infinite set of coupled equations
\begin{equation}
  \frac{1}{\reynolds} \psi_m = \frac{\I k \Psi_0}{4} \frac{\kappa_m - \kappa_{m-1}}{\kappa_m^2 \left(\lambda + \kappa_m^{2}\right)} \psi_{m - 1}
  + \frac{\I k \Psi_0}{4}\frac{\kappa_m - \kappa_{m+1}}{\kappa_m^2 (\lambda + \kappa_m^{2})}\psi_{m + 1}.\label{eq:energy_method_stability_threshold_single_interface}
\end{equation}
We obtain the critical Reynolds number by solving the coupled equations as an
eigenvalue problem numerically in the same fashion as the previous sections. The
results are plotted in wavenumber space in
Figure~\ref{fig:single_boundary_critical_reynolds_k_l}. With both regular
diffusion and hyperdiffusion, the critical Reynolds number increases as $k$ and
$l$ increases. In the limit of $k \ll 1$ and $l \ll 1$, the critical Reynolds
number is zero.

\begin{figure}
  \begin{center}
    \includegraphics[width=2.6in]{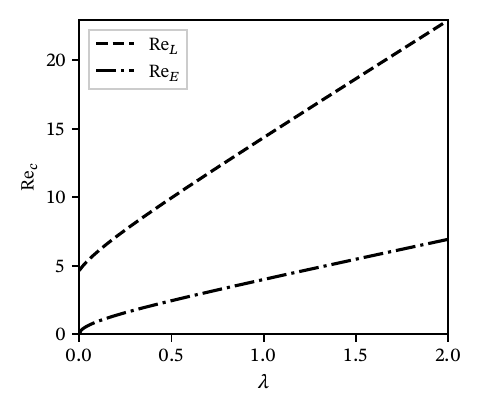}
    \includegraphics[width=2.6in]{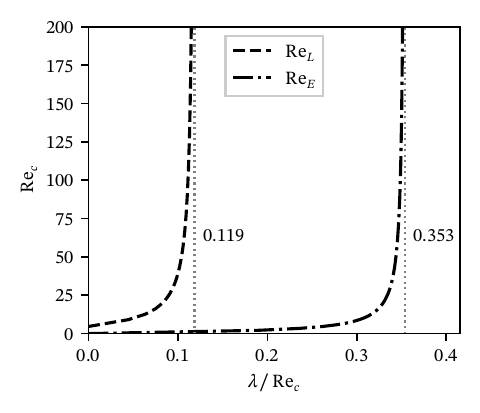}
    \caption{Variation of the critical Reynolds number $\reynolds_c$ with
    damping $\lambda$. Left plot: the upper line is the critical
    Reynolds number for linear instability; the lower line is
    the critical Reynolds number for nonlinear stability (we use the
    notation of Fukuta et al.\ with $\reynolds_L$ for the critical Reynolds
    number calculated by linearisation and $\reynolds_E$ for the critical
    Reynolds number calculated using the energy method). Right plot: the
    critical Reynolds number plotted against the unscaled drag parameter
    $\lambda/\reynolds_c$. The asymptotes shows in the plot correspond to the
    slopes in the first plot as $\lambda \rightarrow \infty$.}
    \label{fig:single_boundary_critical_reynolds_Re_v_lambda}
  \end{center}
\end{figure}


The effect of  damping introduces another parameter to the analysis. As
shown in Figure~\ref{fig:single_boundary_critical_reynolds_Re_v_lambda}, the
critical Reynolds number increases as the  damping is increased. For
comparison, the neutral line computed from linear instability analysis is also
shown in the same plots. Since  damping and diffusion compete to
remove energy from the system, in the presence of damping, a higher
Reynolds number is required to put the system in an unstable state
In the presence of large  damping, its relationship with the linear and
nonlinear critical Reynolds number appears to be linear, as shown in the right
panel. The neutral line from the linear instability analysis, shown as a dashed
line, lies above the line obtained through the energy method. It gives an upper
bound past which the system is unstable. The neutral line obtained
through the energy method gives a lower bound under which the system is
guaranteed to be stable. The system is not guaranteed to be stable in the
absence of damping. The nonlinear critical wavenumber $k_c$ stays at zero even
at $\lambda=100$.




\subsection{Boundary driven finite-depth
fluid}\label{sec:finite-depth-layer-nonlinear-stability}


For a fluid with finite thickness, we use the following Floquet ansatz:
\begin{equation}
  \tilde{\psi} = \e ^{\I kx + \I ly}\sum_{m=-\infty}^{\infty} (\mu_m \cosh\kappa_m z + \nu_m \sinh\kappa_m z) \e ^{\I my}.
\end{equation}
This ansatz takes into account the symmetric and antisymmetric modes of the
system. As in the linear analysis, we study the symmetric and antisymmetric
configuration separately.

For a symmetric configuration, we substitute the ansatz and the symmetric basic
state, \eqref{eq:two_boundaries_basic_theta_symm}, into \eqref{eq:energy_eq}.
The recurrence relations are
\begin{subequations}
  \begin{multline}
    \frac{\mu_m}{\reynolds} =
    - \frac{\I k \Psi_0 \left(\kappa_{m + 1} - \kappa_m\right) \e ^{- z_0} \cosh z_0 \cosh z_0 \kappa_{m + 1}}{2 \kappa_m^2 \left(\lambda + \kappa_m^{2}\right) \cosh z_0 \kappa_m} \mu_{m + 1} \\
    - \frac{\I k \Psi_0 \left(\kappa_{m - 1} - \kappa_m\right) \e ^{- z_0} \cosh z_0 \cosh z_0 \kappa_{m - 1}}{2 \kappa_m^2 \left(\lambda + \kappa_m^{2}\right) \cosh z_0 \kappa_m} \mu_{m - 1}
    \label{eq:energy_method_stability_threshold_double_interface_symm_A}
  \end{multline}
  for the symmetric modes, and
  \begin{multline}
    \frac{\nu_m}{\reynolds} =
    - \frac{\I k \Psi_0 \left(\kappa_{m + 1} - \kappa_m\right) \e ^{- z_0} \sinh z_0 \kappa_{m + 1}  \cosh z_0}{2 \kappa_m^2 \left(\lambda + \kappa_m^{n}\right) \sinh z_0 \kappa_m}  \nu_{m + 1} \\
    - \frac{\I k \Psi_0 \left(\kappa_{m - 1} - \kappa_m\right) \e ^{- z_0} \sinh z_0 \kappa_{m - 1}  \cosh z_0 }{2 \kappa_m^2 \left(\lambda + \kappa_m^{n}\right) \sinh z_0 \kappa_m} \nu_{m - 1}
    \label{eq:energy_method_stability_threshold_double_interface_symm_B}
  \end{multline}
\end{subequations}
for the antisymmetric modes. Likewise, for an antisymmetric configuration, we
substitute the ansatz and the antisymmetric basic state,
\eqref{eq:two_boundaries_basic_theta_antisymm}, into \eqref{eq:energy_eq}. The
recurrence relations are
\begin{subequations}
  \begin{multline}
    \frac{\mu_m}{\reynolds} =
    - \frac{\I k \Psi_0 \left(\kappa_{m + 1} - \kappa_m\right) \e ^{- z_0} \sinh z_0 \sinh z_0 \kappa_{m + 1}}{2 \kappa_m^2 \left(\lambda + \kappa_m^{n}\right) \cosh z_0 \kappa_m}  \nu_{m + 1}\\
    - \frac{\I k \Psi_0 \left(\kappa_{m - 1} - \kappa_m\right) \e ^{- z_0} \sinh z_0 \sinh z_0 \kappa_{m - 1}}{2 \kappa_m^2 \left(\lambda + \kappa_m^{n}\right) \cosh z_0 \kappa_m} \nu_{m - 1}
    \label{eq:eq:energy_method_stability_threshold_double_interface_antisymm_AB}
  \end{multline}
  for the symmetric modes, and
  \begin{multline}
    \frac{\nu_m}{\reynolds} =
    - \frac{\I k \Psi_0 \left(\kappa_{m + 1} - \kappa_m\right) \e ^{- z_0} \sinh z_0 \cosh z_0 \kappa_{m + 1}}{2 \kappa_m^2 \left(\lambda + \kappa_m^{n}\right) \sinh z_0 \kappa_m} \mu_{m + 1} \\
    - \frac{\I k \Psi_0 \left(\kappa_{m - 1} - \kappa_m\right) \e ^{- z_0} \sinh z_0 \cosh z_0 \kappa_{m - 1}}{2  \kappa_m^2\left(\lambda + \kappa_m^{n}\right) \sinh z_0 \kappa_m} \mu_{m - 1}
    \label{eq:eq:energy_method_stability_threshold_double_interface_antisymm_BA}
  \end{multline}
\end{subequations}
for the antisymmetric modes.

\begin{figure}
  \begin{center}
    \includegraphics[width=2.6in]{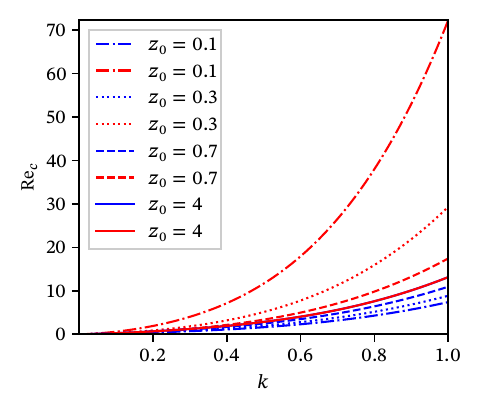}
    \includegraphics[width=2.6in]{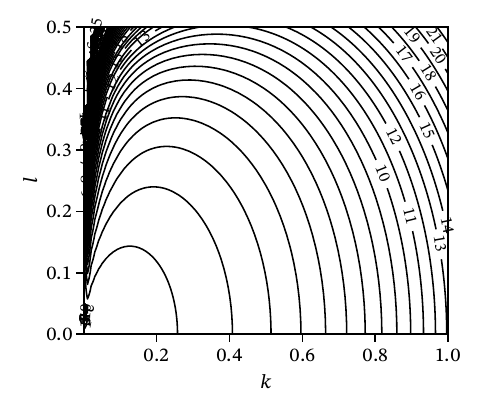}
    \caption{The left panel shows the critical Reynolds number for symmetric
    forcing as a function of $k$ obtained through the energy method at $l=0$.
    Blue (red) curves correspond to a (anti)symmetric modes. The right
    panel shows the critical Reynolds number for symmetric forcing as a
    contour map in wavenumber space.}
    \label{fig:double_boundary_nonlinear_Re_v_k}
  \end{center}
\end{figure}

The critical Reynolds number increases as a function of the horizontal
wavenumber $k$ for both symmetric and antisymmetric modes. The effect of the
separation distance can be seen from the different behaviours between the red
and blue curves in Figure~\ref{fig:double_boundary_nonlinear_Re_v_k}. As the
separation distance $z_0$ is increased, the slope of $\reynolds_c$ with respect
to $k$ decreases for the system for antisymmetric forcing. The opposite
behaviour is true for a system for symmetric forcing. In general, larger
wavenumbers, or smaller length scales, are more stable than smaller wavenumbers
or larger length scales.

\begin{figure}
  \begin{center}
    \includegraphics[width=0.49\textwidth]{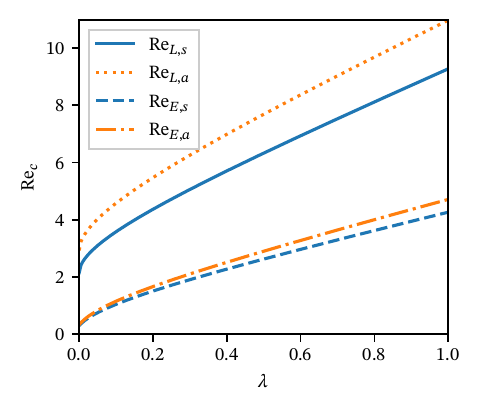}
    \includegraphics[width=0.49\textwidth]{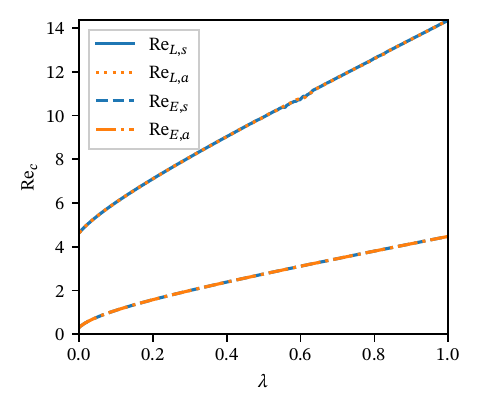}
  \end{center}
  \caption{Critical Reynolds number vs.\ $\lambda$ at two different separating
  distances. The first plot corresponds to $z_0=1.5$. The second plot
  corresponds to $z_0=10$.}
  \label{fig:energy_method_re_c_v_lambda}
\end{figure}

\begin{figure}
  \begin{center}
    \includegraphics[width=0.6\textwidth]{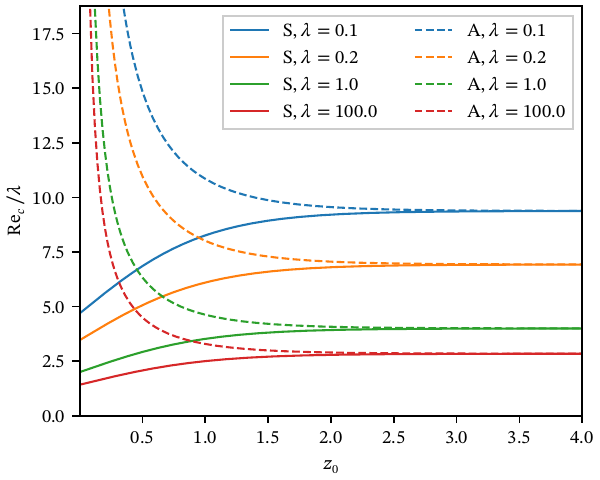}
  \end{center}
  \caption{Critical Reynolds number vs.\ $z_0$ using the energy method.
  Solutions form the (anti)symmetric configuration are plotted in (dashed) solid
  curves.}
  \label{fig:energy_method_re_c_v_z0}
\end{figure}

The critical Reynolds number as a function of damping is shown in
Figure~\ref{fig:energy_method_re_c_v_lambda}. A system with antisymmetric
forcing is more stable, linearly or nonlinearly, than a system with symmetric
forcing. As the separation distance is increased, the neutral curves approach
that of a semi-infinite SQG+ system, shown in
\ref{fig:single_boundary_critical_reynolds_Re_v_lambda}. Both the linear and
nonlinear critical Reynolds number appear to be linear functions at sufficiently
large damping. $\reynolds_c/\lambda$ is plotted against $z_0$ in
Figure~\ref{fig:energy_method_re_c_v_z0}. Both the symmetric and antisymmetric
curves converge to horizontal lines at sufficiently large $z_0$. The levels of
these horizontal lines correspond to values in a semi-infinite SQG system.








\section{SQG+ Linear Instability of a Semi-Infinite Layer}\label{sec:linear-instability-order-rossby}

Studying all the configurations covered in the previous sections with order
Rossby corrections using linearisation and the energy method presents numerous
challenges. The study of linear instability of a layer forced on two boundaries
is technically possible, but the conventional approach of finding a recurrence
relation and solving with an eigenvalue solver cannot be directly applied here.
One can also show that studying SQG+ nonlinear stability even for the simpler
configuration of a semi-infinite layer using the correction terms derived by
\cite{muraki1999} and \cite{hakim2002} is infeasible. For these practical
considerations, for the remainder of the paper, we limit the ourselves to
studying only the semi-infinite layer in SQG+.

With order Rossby corrections, the existing notation used in the previous
sections is no longer adequate. In SQG+, a source vector field replaces the
streamfunction~\citep{muraki1999}. Using the $\Phi$ component of the vector
field as an example, we rewrite the component as a perturbation expansion in the
Rossby number $\rossby \equiv U/fL$,
\begin{equation}\label{eq:phi_rossby_expansion}
  \Phi(x,y,z,t) = \Phi^0(x,y,z,t) + \rossby \Phi^1(x,y,z,t) + \mathcal{O}(\rossby^2).
\end{equation}
In the $\rossby \ll 1$ limit, we recover $\Phi=\Phi^0$ as the SQG
streamfunction. To study the linear instability, this expansion is further
expanded to account for small perturbations around a periodic basic state,
\begin{subequations}
  \label{eq:phi_disturbance_expansion}
  \begin{align}
    \Phi^0(x,y,z,t) &= \Phi^{00}(y,z) + \disturb \Phi^{01}(x,y,z,t) + \mathcal{O}(\disturb^2)\label{eq:stream_func_disturbance_o1} \\
    \Phi^1(x,y,z,t) &= \Phi^{10}(y,z) + \disturb \Phi^{11}(x,y,z,t) + \mathcal{O}(\disturb^2),\label{eq:stream_func_disturbance_o_rossby}
  \end{align}
\end{subequations}
where, on the right hand side, the first digit in the superscript indicates the
order in the Rossby number, and the second digit in the superscript indicates
the order in the small disturbance expansion, with $\disturb$ the magnitude
of this disturbance. Combining the above, the $\Phi$ component of the vector field
can be written as
\begin{equation}
  \Phi(x,y,z,t) = \Phi^{00}(y,z) + \rossby \Phi^{10}(y,z) + \disturb \Phi^{01}(x,y,z,t) + \rossby\disturb \Phi^{11}(x,y,z,t) + \ldots.
\end{equation}
We will follow this procedure in the following passage, expanding first in the
Rossby number followed by an expansion in the small disturbance limit.

Using the notation we have established, we expand the advection
equation~\eqref{eq:fullAdvEq} in the Rossby number
using~\eqref{eq:phi_rossby_expansion}, limiting ourselves to terms up to
${O}(\rossby)$,
\begin{equation}
  \label{eq:rossby_expanded_time_evo}
  \dt{}{}\left(\Phi^0_z + \rossby \Phi^1_z\right) + (\vec{u}^0 + \rossby \vec{u}^1)\dotprod\nabla (\Phi^0_z + \rossby \Phi^1_z)
  = -\frac{1}{\reynolds}\{(-1)^n\laplace^n+ \lambda\}(\Phi^0_z + \rossby \Phi^1_z) + F.
\end{equation}
With the explicit next order correction terms derived by \cite{muraki1999} and
\cite{hakim2002}, we rewrite \eqref{eq:rossby_expanded_time_evo} in terms of
$\Phi^0$ and various tilde terms that will be expanded in later steps,
\begin{subequations}
  \label{eq:velocity_rossby_expansion}
  \begin{multline}
    \Phi^0_{zt} + \rossby (\Phi^0_z \Phi^0_{zzt} + \Phi^0_{zz} \Phi^0_{zt} + \tilde{\Phi}^1_{zt})
    + u (\Phi^0_{xz} + \rossby (\Phi^0_z \Phi^0_{xzz} + \Phi^0_{zz} \Phi^0_{xz} + \tilde{\Phi}^1_{xz})) \\
    + v (\Phi^0_{yz} + \rossby (\Phi^0_z \Phi^0_{yzz} + \Phi^0_{zz} \Phi^0_{yz} + \tilde{\Phi}^1_{yz})) \\
    = - \frac{(-1)^{n}}{\reynolds} \laplace^n(\Phi^0_z + (\rossby (\Phi^0_z \Phi^0_{zz} + \tilde{\Phi}^0_z))
    + \frac{\lambda}{\reynolds} (\Phi^0_z + \rossby (\Phi^0_z \Phi^0_{zz} + \tilde{\Phi}^0_z)),
  \end{multline}
  where
  \begin{align}
    u &= -\Phi^0_y
      - \rossby \Phi^0_y\Phi^0_{zz}
      - 2\rossby \Phi^0_z\Phi^0_{yz}
      - \rossby \tilde{\Phi}^1_y
      - \rossby \tilde{F}^1_z \\
    v &= \Phi^0_x
      + \rossby \Phi^0_x\Phi^0_{zz}
      + 2\rossby \Phi^0_{xz}\Phi^0_z
      + \rossby \tilde{\Phi}^1_x
      - \rossby \tilde{G}^1_z.
  \end{align}
\end{subequations}

Next, we expand using~\eqref{eq:phi_disturbance_expansion}.
The expansion is too long to be shown in full here. After the expansion, the velocity terms
\eqref{eq:velocity_rossby_expansion} are, up to ${O}(\rossby)$,
${O}(\disturb)$, and ${O}(\disturb\rossby)$, with
$\kappa=\sqrt{k^2+l^2}$,
\begin{subequations}
  \label{eq:vel_expansion}
  \begin{multline}
    \label{eq:u_expansion}
    u = - \Phi^{00}_y
      - \rossby \Phi^{00}_y \Phi^{00}_{zz}
      - 2\rossby \Phi^{00}_z\Phi^{00}_{yz}
      - \disturb \Phi^{01}_y
      - \disturb \rossby \Phi^{00}_{zz}\Phi^{01}_y
      - 2\disturb \rossby \Phi^{00}_{yz} \Phi^{01}_z
      - \disturb \rossby \Phi^{00}_y \Phi^{01}_{zz} \\
      - 2\disturb \rossby \Phi^{00}_z \Phi^{01}_{yz}
      - \mathcal{F}^{-1}\left\{
        -\frac{e^{-|\kappa|z}}{|\kappa|}\mathcal{F}\left[
          \rossby\tilde{\Phi}^1_{yz}\Big|_{z=0}
        \right]
      \right\}
      - \mathcal{F}^{-1}\left\{
        -|\kappa|e^{-|\kappa|z}\mathcal{F}\left[
          \rossby\tilde{F}^1\Big|_{z=0}
        \right]
      \right\}
  \end{multline}
  \begin{multline}
    \label{eq:v_expansion}
    v = \disturb \Phi^{01}_x
      + \disturb \rossby \Phi^{00}_{zz}\Phi^{01}_x
      + 2\disturb \rossby \Phi^{00}_{z}\Phi^{01}_{xz}
      - \mathcal{F}^{-1}\left\{
        -|\kappa|e^{-|\kappa|z}\mathcal{F}\left[
          \rossby\tilde{G}^1\Big|_{z=0}
        \right]
      \right\} \\
      + \mathcal{F}^{-1}\left\{
        -\frac{e^{-|\kappa|z}}{|\kappa|}\mathcal{F}\left[
          \rossby\tilde{\Phi}^1_{xz}\Big|_{z=0}
        \right]
      \right\},
  \end{multline}
  where
  \begin{align}
    \tilde{F}^1 &= -\mathcal{F}^{-1}\left\{e^{-|\kappa|z}\mathcal{F}\left[\left(\Phi^{00}_y\Phi^{00}_z+\disturb\Phi^{00}_z\Phi^{01}_y+\disturb\Phi^{00}_y\Phi^{01}_z\right)\Big|_{z=0}\right]\right\} \\
    \tilde{G}^1 &= \disturb\mathcal{F}^{-1}\left\{e^{-|\kappa|z}\mathcal{F}\left[\left(\Phi^{00}_z\Phi^{01}_x\right)\Big|_{z=0}\right]\right\} \\
    \tilde{\Phi}^1 &= -\mathcal{F}^{-1}\left\{-\frac{e^{-|\kappa|z}}{|\kappa|}\mathcal{F}\left[\left(\Phi^{00}_z\Phi^{00}_{zz}+\disturb\Phi^{01}_z\Phi^{00}_{zz}+\disturb\Phi^{01}_{zz}\Phi^{00}_z\right)\Big|_{z=0}\right]\right\}.
  \end{align}
\end{subequations}
At this stage, all ${O}(\rossby)$ terms have been rewritten in terms of
the SQG streamfunction using the diagnostic expressions derived in
\cite{hakim2002}. Every component in the full expansion is now solely a function
of $\Phi^{00}$ and $\Phi^{01}$.

Using the notation of this section, the basic state is
\begin{subequations}
  \begin{equation}
    \Phi^{00} = \Psi_0 e^{z} \sin y.
  \end{equation}
  Treating the problem as a Floquet problem, we define
  \begin{equation}
    \Phi^{01} = e^{\sigma t+\I kx+\I ly+\kappa z} \sum_{m=-\infty}^{\infty}\phi_m e^{\I my}.
  \end{equation}
\end{subequations}
Evaluating the expanded equation at $z=0$, the linear growth rate of an SQG+
Kolmogorov system in a semi-infinite layer is governed by a recurrence relation,
\begin{multline}
  \label{eq:single_interface_recurrence_relation}
  \sigma {\phi}_{m} = - \frac{\rossby k \Psi_0^{2} \left({\kappa}_{m + 1} - {\kappa}_{m + 2}\right) \left({\kappa}_{m + 1} - {\kappa}_{m + 2} - 1\right) {\phi}_{m + 2}}{4 {\kappa}_{m + 1} {\kappa}_{m}}
  + \frac{\I k \Psi_0 \left({\kappa}_{m + 1} - 1\right) {\phi}_{m + 1}}{2 {\kappa}_{m}} \\
  + \frac{\rossby k \Psi_0^{2} \left({\kappa}_{m - 1} - {\kappa}_{m - 2}\right) \left({\kappa}_{m - 1} - {\kappa}_{m - 2} - 1\right) {\phi}_{m - 2}}{4 {\kappa}_{m - 1} {\kappa}_{m}}
  + \frac{\I k \Psi_0 \left({\kappa}_{m - 1} - 1\right) {\phi}_{m - 1}}{2 {\kappa}_{m}} \\
  + \left(\frac{\rossby k \Psi_0^{2} \left({\kappa}_{m + 1} - {\kappa}_{m - 1}\right) \left({\kappa}_{m + 1} {\kappa}_{m - 1} - {\kappa}_{m}^{2} - {\kappa}_{m}\right)}{4 {\kappa}_{m + 1} {\kappa}_{m - 1} {\kappa}_{m}} - \frac{\lambda + {\kappa}_{m}^{2 n}}{\reynolds}\right) {\phi}_{m},
\end{multline}
where $\kappa_m^2 = k^2 + (l + m) ^2$ for clarity. With $\rossby=0$, this
reduces to the SQG recurrence relation
\eqref{eq:single_interface_o1_recurrence_relation}.

\begin{figure}
  \begin{center}
    \includegraphics[width=\textwidth]{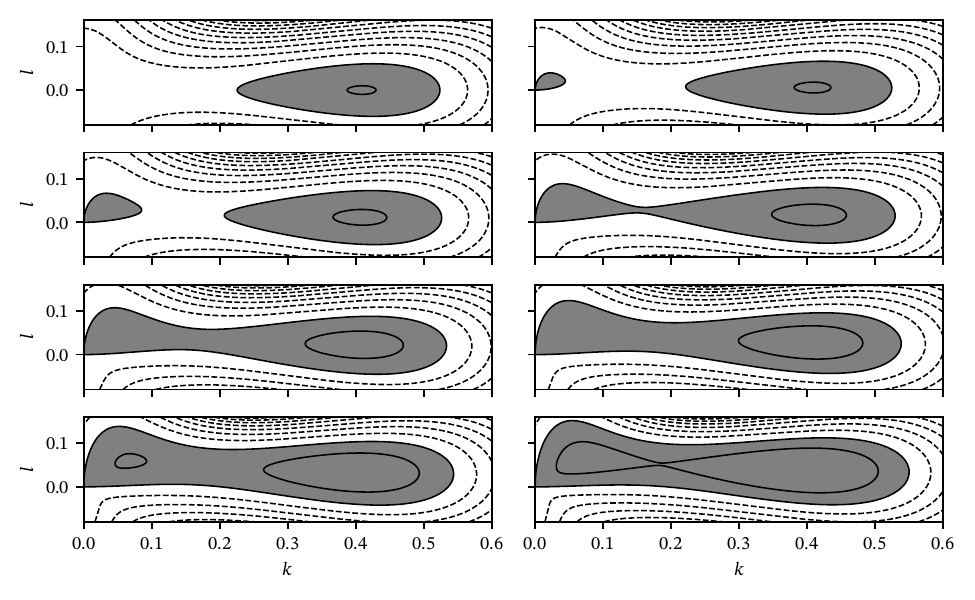}
    \caption{Growth rate diagrams for an SQG+ semi-infinite layer at various
    Rossby numbers. $\reynolds=5$. Left to right, top to bottom: $\rossby=0$,
    $0.0571$, $0.1143$, $0.1714$, $0.2286$, $0.2857$, $0.3429$ and $0.4$, with
    with $n=1$ and $\lambda=0$. The coupled equations are
    truncated at $N=16$. The spacing between levels is $0.004$.}
    \label{fig:sqg_plus_contour_var_rossby}
  \end{center}
\end{figure}

\begin{figure}
 \begin{center}
   \includegraphics[width=0.48\textwidth]{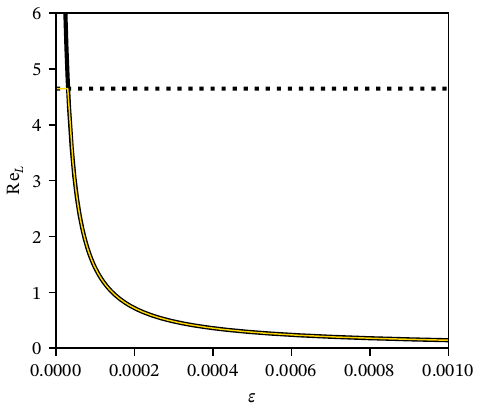}
   \includegraphics[width=0.48\textwidth]{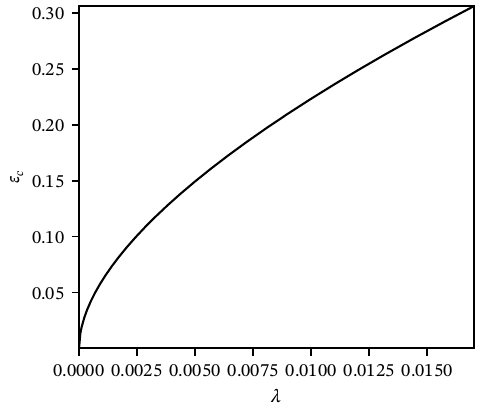}
   \caption{Left:  critical Reynolds number vs.~$\rossby$. The dashed black
   curve is the critical Reynolds number of the region of instability around
   $k=0.36$. The solid black curve is the critical Reynolds number of the region
   of instability around $k=0$. The two curves intersect at
   $\rossby=\num{3.0957e-5}$. The thin yellow curve in the first plot shows the
   effective critical Reynolds number as a function of $\rossby$. The
   intersection's relationship with damping is shown in the second plot.}
   \label{fig:sqg_plus_critical_reynolds_vs_rossby}
 \end{center}
\end{figure}

Figure~\ref{fig:sqg_plus_contour_var_rossby} shows the changes of the linear
growth rate diagrams of a semi-infinite system with increasing Rossby number.
The Reynolds number is chosen to be just above the SQG critical value. The
introduction of ageostrophic corrections destroys the isotropy in wavenumber
space. With $\rossby \ne 0$, the regions of instability in the $k > 0$
half-plane are no longer symmetric about $l=0$, but skews towards $l>0$.
Furthermore, as $\rossby$ increases, a new region of instability develops near
$(k, l) = (0, 0)$. As the Rossby number increases, the regions of instability
visibly grow and they merge when $0.11 < \rossby < 0.17$. Simultaneously, the
region of instability near $(k,l)=(0,0)$ becomes more unstable. At
$\rossby=\num{3.0957e-5}$, this region of instability overtakes the region to
its right to become the main destabilising modes of the system. The
destabilising effect of the region of instability near $(k, l) = (0, 0)$ is
illustrated in the left panel of
Figure~\ref{fig:sqg_plus_critical_reynolds_vs_rossby}. With zero damping and an
increasing Rossby number, the critical Reynolds number drops rapidly at
$\rossby=\num{3.0957e-5}$. By $\rossby=0.001$, the linear critical Reynolds
number of an SQG+ system is an order of magnitude lower than that of an SQG
system. This effective critical Reynolds number is represented graphically by
the thin yellow curve in the same plot. In other words, in the presence of
ageostrophic effects, an SQG+ system is linearly unstable even with very strong
viscous dissipation due to this long-wave instability.

\begin{figure}
  \begin{center}
    \includegraphics[width=\textwidth]{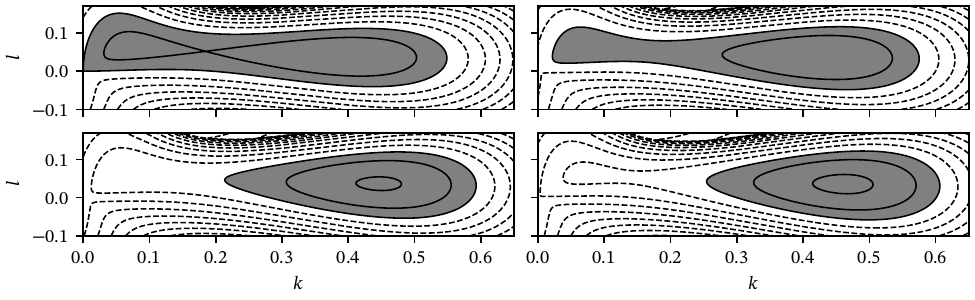}
  \end{center}
  \caption{Growth rate diagrams for an SQG+ semi-infinite layer at
  $\rossby=0.4$. The damping coefficient is, from left to right, top to bottom,
  $0$, $0.0167$, $0.033$, and $0.05$. These plots are all plotted at
  $1.1\reynolds_c(\rossby, \lambda)$ computed from the black, dotted curve in
  Figure~\ref{fig:sqg_plus_critical_reynolds_vs_rossby}, which are, in the same
  order, $4.530$, $4.837$, $5.089$, and $5.318$.}
  \label{fig:sqg_plus_lambda_vs_rossby_contour}
\end{figure}

\begin{figure}
  \begin{center}
    \includegraphics[width=\textwidth]{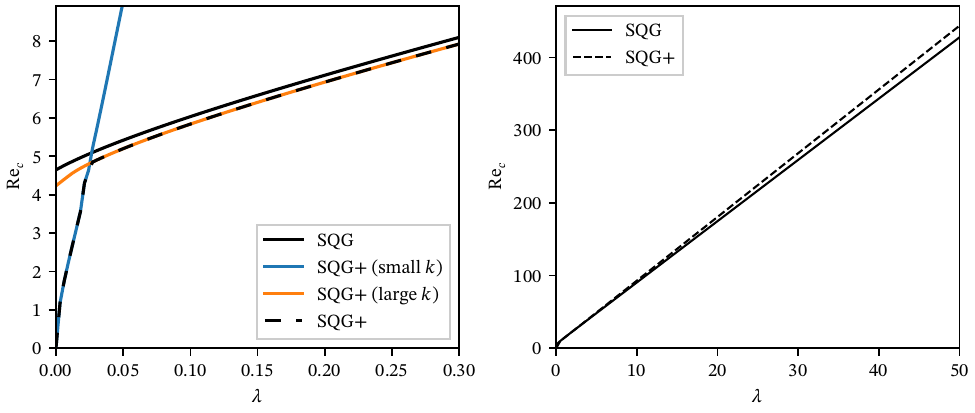}
  \end{center}
  \caption{Linear critical Reynolds number plotted as a function of the damping
  parameter $\lambda$. The Rossby number is set to $0.4$. As shown in
  Figure~\ref{fig:sqg_plus_contour_var_rossby}, there are two disconnected
  regions of instability in wavenumber space. The orange curve shows $\reynolds_c$
  in $k>0.3$, and the blue curve shows $\reynolds_c$ for the region the long
  length scale limit (small $k$). The black dashed curve shows the effective
  critical Reynolds number. The curve representing the SQG critical Reynolds
  number is shown for comparison.
  }
  \label{fig:sqg_plus_re_c_v_lambda}
\end{figure}

\begin{figure}
  \begin{center}
    \includegraphics[width=0.5\textwidth]{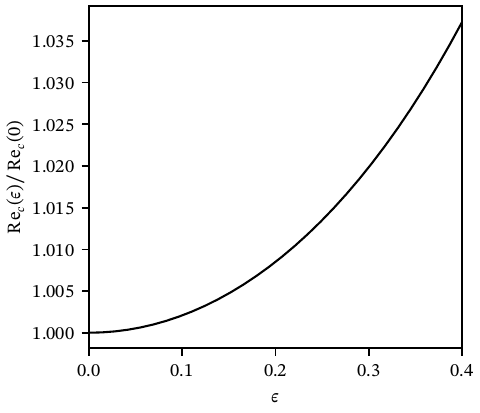}
  \end{center}
  \caption{The slope of the SQG+ curve in Figure~\ref{fig:sqg_plus_re_c_v_lambda}
  as a function of the Rossby number at $\lambda=50$. The results are
  normalised against the SQG result.}
  \label{fig:sqg_plus_re_c_over_lambda_v_rossby}
\end{figure}

The destabilising effect of ageostrophic corrections is easily counteracted by
damping. The critical Rossby number at which this long-wave instability
destablises the system increases with the presence of damping and is shown as a
function of $\lambda$ in the right panel of
Figure~\ref{fig:sqg_plus_critical_reynolds_vs_rossby}. In other words, the
presence of small damping pushes the ``cliff'' in the critical Reynolds number
shown in the left panel to much higher Rossby numbers.
Figure~\ref{fig:sqg_plus_lambda_vs_rossby_contour} shows the growth rate
diagrams with varying $\reynolds_c$ and $\lambda$ at a fixed $\rossby=0.4$. With
increasing $\lambda$, the region of instability near $(k,l)=(0,0)$ is
suppressed. While the growth rate diagrams are still anisotropic in wavenumber
space, the regions of stability are approaching the teardrop shape of that of an
SQG growth diagram. The stabilising power of damping is further demonstrated in
Figure~\ref{fig:sqg_plus_re_c_v_lambda}. The blue curve shows $\reynolds_c$
associated with the region of instability near the origin of the wavenumber
space, whereas the orange curve shows $\reynolds_c$ of the region of instability to
the right. As $\lambda$ increases, the critical Reynolds number of the region of
instability quickly increases, which indicates this region of stability is
becoming more stable. By $\lambda=0.025$, it is overtaken by the other centre of
instability as the main destabilising mode.

The equivalent SQG curve is superposed on this graph. With small damping, an SQG
system is more linearly stable than an SQG+ system. However, an SQG+ system
stabilises more readily with increasing damping than an SQG system, and an SQG+
system is the more stable of the two with sufficiently large damping.
Figure~\ref{fig:sqg_plus_re_c_over_lambda_v_rossby} shows the ratio of the
critical Reynolds number for SQG+ over SQG systems with varying $\rossby$ at
large $\lambda$.

\section{Discussion}\label{sec:discussion}


In this paper, we have studied the linear and nonlinear stability of various SQG
and SQG+ configurations using linearisation and the energy method. We have shown
that, in contrast to a two-dimensional system, the initial linear
destabilisation of a semi-infinite SQG system does not originate from the long
wave limit. On the other hand, nonlinear analysis shows that $k=0$ remains the
least stable mode, even with very strong damping. $\reynolds_{Ec}$ is $0$ in the
absence of damping. With the increase of damping, the system becomes more
stable. With sufficiently large damping, both the linear and nonlinear critical
Reynolds number grow linearly with damping.

When an SQG system in a fluid layer of finite thickness is being forced from
both the upper and lower boundaries, its behaviour is in between that of a
semi-infinite SQG system and a 2D system: in a fluid layer with large enough
thickness, the behaviour tends toward that of a semi-infinite SQG system, while
for small thickness the behaviour of a symmetric configuration tends toward that
of a 2D system. However, the antisymmetric configuration deviates from this
pattern and becomes less prone to instability as the layer thickness is
decreased, since the effects of the forcing on the two boundaries counteract one
another in the fluid interior. Its nonlinear critical Reynolds number approaches
infinity at the $z_0 \ll 1$ limit. In the small depth limit, the hyperbolic sine
function in \eqref{eq:two_boundaries_basic_psi_antisymm} approaches zero, and
the boundary forcing essentially vanishes.

For a semi-infinite SQG+ system, we have also studied the impact of the
ageostrophic corrections on stability. We found that in the absence of damping,
the ageostrophic corrections greatly destabilise the system through
instabilities in the long wave limit. The critical Reynolds number decreases
rapidly as a region of instability develops near $(k,l)=(0,0)$. We also found
that this instability is very effectively tamed by the introduction of damping.
Due to the enlargement of the regions of instability, an SQG+ system is more
unstable than an equivalent SQG system when damping is weak. However, in the
large damping regime, the opposite is true. We do not currently have a
satisfactory explanation for this phenomenon.

While SQG is only a model for the ocean's submesoscale, the results of this
paper suggest that the stability properties of an SQG system can be physically
different from its 2D Euler counterpart. The dynamics will be dominated by
physical processes that are noticeably smaller than the system length scale,
unlike 2D Euler flows. The dynamics of an SQG system with two active layers
depends on the thickness on the distance between the layers, a model for the
oceanic mixed layer. A mixed layer that is thicker behaves more like a
semi-infinite SQG system and a mixed layer that is sufficiently thin behaves
more like a 2D Euler system. Our results with order Rossby corrections suggest
that, under the influence of ageostrophic effects, instabilities originate from
the eddies at the system length scale, more like a 2D system. However, the
presence of drag shifts the dominant processes from the system length scale down
to a smaller length scale. Between drag and ageostrophic effects lies rich
physics that is worth exploring in future studies.


\section*{Acknowledgements}

The authors would like to acknowledge inspiring discussions with
W.~R.~Young.

\section*{CRediT}
\textbf{Mac Lee}: Software, Formal analysis, Investigation, Writing - Original
Draft \textbf{Stefan Llewellyn Smith}: Conceptualization, Methodology,
Validation, Writing - Review \& Editing, Supervision, Funding acquisition

\section*{Funding}

This research was partially funded by an ONR NISEC award through UCSD.

\section*{Declaration of interests}

The authors report no conflict of interest.


\bibliographystyle{elsarticle-num}
\bibliography{references}
\end{document}